\documentclass{emulateapj} 

\catcode`\"=\active\let"=\"  

\def\3{{\ss} }

\def\c12{{1\over 2}}

\def\plusplus{\raise 0.3ex\hbox{${\scriptstyle ++}$}{}}

\shorttitle{Signatures of tides in dSph galaxies}  
\shortauthors{Pe\~{n}arrubia et al.}  
\slugcomment{Submitted to ApJ}  
  
\newcommand{\oversim}[2]{\protect{\mbox{\lower0.5ex\vbox{%
   \baselineskip=0pt\lineskip=0.2ex   
   \ialign{$\mathsurround=0pt #1\hfil##\hfil$\crcr#2\crcr\sim\crcr}}}}}    
\newcommand{\simless} {\mbox{$\,\mathrel{\mathpalette\oversim<}\,$}} 
\begin{document}   
   
\title[The Signature of Galactic Tides in dSphs]{The Signature of Galactic Tides in Local Group Dwarf Spheroidals}
   
\author{Jorge Pe\~{n}arrubia\altaffilmark{1,2}, Julio F. Navarro\altaffilmark{2,3},   
Alan W. McConnachie\altaffilmark{2,4} \& Nicolas
F. Martin\altaffilmark{5} }

\email{jorpega@ast.cam.ac.uk}   
\altaffiltext{1}{Institute of Astronomy, University of Cambridge, Madingley Road, Cambridge CB3 0HA, UK}
\altaffiltext{2}{Department of Physics and Astronomy, University of Victoria,   
3800 Finnerty Rd., Victoria, BC, V8P 5C2, Canada} 
\altaffiltext{3}{Fellow of the Canadian Institute for Advanced Research}   
\altaffiltext{4}{Herzberg Institute of Astrophysics, 5071 West Saanich Road, Victoria, BC V9E 2E7, Canada}
\altaffiltext{5}{Max-Planck-Institut fuer Astronomie, K\"onigstuhl 17, D-69117 Heidelberg, Germany}
\begin{abstract}   
We use N-body simulations to explore the effects of tidal stripping on
the structure of dwarf spheroidal galaxies (dSphs). Our models assume
cosmologically motivated initial conditions where dSphs are modeled as
King spheres, embedded in Navarro-Frenk-White dark halos and orbiting
the Galactic potential on eccentric orbits. As expected, systems that
orbit through dense regions of the Galaxy lose a significant fraction
of stars after each pericentric passage. These episodes of mass loss,
however, do {\it not} impose a clear tidal cutoff on the bound stellar
core. Rather, once equilibrium has been re-established the outer mass
profile approaches a power-law well described by a simple Plummer
model. As noted in earlier work, tides also result in transient
features in the outer density profile. As the system relaxes, an
outward-moving ``excess'' of stars is found at radii where the local
crossing time exceeds the time elapsed since pericenter. If the orbit
of the dSph is known, these results provide a simple way to assess
whether ``breaks'' and ``bumps'' in the outer profile of dSphs are
actually tidal in origin. We apply this to the Sagittarius dwarf and,
encouragingly, identify two features in the surface brightness profile
that may be traced to its two last pericentric passages. Applied to
Leo I, our results predict that any tidal break would occur at radii
beyond those surveyed by current data, casting doubt on recent claims
of the detection of tidal debris around this galaxy. For Carina, our
model indicates that the tidal break should occur at a radius twice
farther than observed. This suggests that the outer excess of stars in
Carina is not tidal in origin unless its orbit is in error. A similar
comment applies to Sculptor, whose pericenter appears too large for
Galactic tides to be important but whose outer profile, like that of
Draco, nonetheless follows closely a Plummer-law. Fornax and Leo II
show no sign of a power-law outer profile, suggesting that they have
not been significantly affected by tides. Published profiles for other
Milky Way dSph companions do not extend sufficiently far to allow for
conclusive assessment. Panoramic surveys that extend surface
brightness profiles beyond $\sim 10$ core radii, together with
improved constraints on the orbital parameters of dSphs, are needed in
order to establish the true origin of the outer envelopes of stars
surrounding dSphs.
\end{abstract}   
    
\section{Introduction}\label{sec:intro}

The last decade has seen the emergence of a paradigm for the matter
content and geometry of the Universe, where structures form as gravity
amplifies primordial fluctuations in the dominant form of matter in
the Universe: Cold Dark Matter (CDM). One virtue of the CDM paradigm
is that, once its parameters are fixed by observations of large scale
structures (Mpc and larger), it can be used to make precise
predictions on small scales. Verifiable predictions on the scale of
galaxies are particularly important, given the numerous observations
that constrain the content and spatial distribution of dark matter in
galaxies of various types.

The most discerning tests apply to faint galaxies. This is because the
luminous (baryonic) component of galaxies is of lesser dynamical
importance in these systems (the most dark-matter dominated galaxies
known are dwarfs), and dynamical constraints may therefore be compared
directly with CDM predictions even when full understanding of the
galaxy formation mechanism may be lacking. Further, the properties of
galaxies and dark matter diverge on small scales: we expect many more
small dark matter halos than there are known faint galaxies (White \&
Rees 1978, Kauffmann et al 1993, Cole et al 1994). Understanding how
one population maps onto the other would shed light on the nature of
dark matter and on the main mechanisms responsible for galaxy
formation.

This line of inquiry is potentially most fruitful when applied to the
faintest galaxies known, since their properties are expected to be
extraordinarily sensitive to the detailed ingredients of galaxy
formation models and to the nature of dark matter.  The dark matter
content of dwarf spheroidal galaxies (dSphs) is thus an issue of
crucial importance in galaxy formation studies. 

Because they are extremely faint, dSphs can only be studied in the
Local Group, where they appear to cluster tighly around M31 and the
Milky Way. These systems are therefore thought to inhabit the small
systems making up the abundant ``substructure'' seen in cosmological
simulations of CDM halos (Klypin et al 1999, Moore et al 1999).

The sheer number of these substructure halos (``subhalos'', for short)
poses an interesting challenge: the most recent simulations indicate
that the Milky Way halo may host thousands of subhalos massive enough,
in principle, to host faint galaxies, a number that far exceeds the
known luminous Galactic satellites. One way of reconciling this
discrepancy is to assume that luminous satellites inhabit only the
(few) most massive subhalos (Stoehr et al 2002). Alternatively, the
scarcity of luminous satellites may reflect the operation of a
physical mechanism that ``lights up'' a small selection of subhalos
with a wide range of masses depending, for example, on their collapse
time, density, and/or environment (Bullock et al 2002, Benson et al
2002, Madau et al 2008). Strong constraints on the total mass and
radial extent of the dark halos surrounding dSphs would help
distinguish between these possibilities.

Stars are the only viable dynamical tracer in dSphs, so direct mass
estimates rely on the kinematics and radial extent of their stellar
component. This yields constraints on the total mass within the
luminous radius of each dwarf but provides little information on the
extent or mass of a surrounding halo (Pe\~narrubia, McConnachie \&
Navarro 2008, hereafter Paper I). Probes
of the outer halos of dSphs are, therefore, necessarily indirect, but
crucial to elucidate the cosmological origin of these systems.

The response of satellites to the Galactic tidal field provides a
promising way to make progress. Ever since dSphs were discovered (Hubble 1936, Holmberg 1950), Galactic tides have been thought to play a significant role
in sculpting their structure. This assumption finds inspiration in the
fact that the {\it shape} of the surface brightness profiles of dSphs
resembles that of globular clusters, with a well-defined central
region of nearly constant stellar density (a ``core'') and a sharp
decline in the outer parts (e.g. Irwin \& Hatzidimitriou 1995). 
King (1962) profiles capture
these features well, and have traditionally been the parametrization
of choice for dSph profiles. The outer cutoff in the luminous profile
has therefore often been identified with a ``tidal radius'' imposed by
the Galactic potential, just like in King (1966) models of globular
clusters. 

This interpretation is, however, debatable, given that dSphs are
substantially more massive than globulars and orbit mostly the outer
Galactic halo on long period orbits. It is thus still unclear whether
the outer limit in the stellar distribution of a dSph has true
dynamical significance or whether it just reflects a sharp boundary in
the region where star formation was able to proceed efficiently.

On the other hand, if Galactic tides {\it are} actively stripping
stars from dSphs, it would be clear that these systems are unlikely to
be surrounded at present by massive dark halos extending far beyond
their luminous radius. This is because the outer parts of such halos
would have been stripped long before the stars (see, e.g.,
Pe\~narrubia, Navarro \& McConnachie 2008, hereafter Paper II),
implying that the total self-bound mass associated with a dwarf does
not exceed greatly the dynamical mass estimated from its stellar size
and kinematics. Dynamical masses have been measured for most dwarfs,
and, intriguingly, seem almost independent of dSph luminosity (see,
e.g., the review of Mateo 1998, Simon \& Geha 2007, Walker et
al. 2007, Gilmore et al. 2007, Martin et al. 2007 and also our Paper I). 
This property
seems to be shared by the ultra-faint dSph population recently
discovered around the Milky Way (e.g Zucker et al. 2004, 2006a,b,c,
Willman et al. 2005, Belokurov et al. 2006a, 2007, 2008; 
Irwin et al. 2007). Strigari et al (2008), for example,
conclude that dSphs spanning almost 7 decades in luminosity share a
common $\sim 10^7 M_{\odot}$ mass scale within a 300 pc radius. It
would be interesting to see whether this result also holds for the
dSph population recently discovered around M31 (Zucker et al. 2004,
2007; Martin et al. 2006; Majewski et al. 2007; Ibata et al. 2007;
Irwin et al. 2008; McConnachie et al. 2008).

If this is truly the total mass associated with dSphs, then strong
constraints on how these galaxies formed would follow. Cosmological
simulations indicate that one may expect roughly {\it one thousand}
$\sim 10^7 M_{\odot}$ CDM subhalos around the Milky Way (Diemand et al
2008, Springel et al 2008), challenging models to explain what led
only a tiny fraction of such systems to host luminous
dwarfs. Confirming whether dSphs are actually losing stars to Galactic
tides, and therefore inhabit low-mass subhalos, would thus provide a
really important clue to the origin of these systems.

We examine these issues here with the aid of N-body simulations of
cosmologically-motivated dSph models, where the stellar component is
represented by King (1966) spheres embedded within Navarro, Frenk \&
White (1996, 1997, hereafter NFW) dark matter halos. Our main goal is
to identify unambiguous signatures of the effect of Galactic tides on
the stellar structure of a dwarf spheroidal.  In particular, we
examine the effect of tidal mass loss on the surface brightness
profile, as well as the significance of transient ``bumps'' and
``breaks'' in the outer profile caused by tides. Such features have
been reported in the literature for a number of dwarfs, but it is
still unclear how or whether they relate to the effect of Galactic
tides.

The plan for this paper is as follows.  In Sec. 2 we introduce our
numerical models, and outline the main results in Sec. 3. We apply
these results in Sec. 4 to individual Milky Way dSph
satellites. Sec. 5 summarizes our main findings.

\section{Numerical simulations}
\label{sec:numsim}

Our numerical setup has been presented in detail in Paper II and we
refer the interested reader to that contribution for details. For
completeness, and in order to introduce useful notation, we describe
here the main aspects of the modeling procedure.

\subsection{The host galaxy}\label{sec:hostmodel}

We model the host galaxy potential with an NFW profile, given by
\begin{equation}  
\Phi_{\rm NFW}(r)=-V_{\rm max}^2 \, {r_{\rm max}\over r}\, {\ln(1+r/r_s)},  
\label{eq:phi}  
\end{equation}  
where $V_{\rm max}$ and $r_{\rm max}$ identify the peak of the
circular velocity profile. The radial scale $r_{\rm max}$ is $2.16\,
r_s$, the usual scale radius of the NFW profile. The host potential is
assumed fixed during the evolution of the dwarf. Since our main goal is to identify the signatures of Galactic tidal stripping in dwarfs, for simplicity we shall neglect in this contribution the effects introduced by subhalo-subhalo interactions\footnote{We refer interested readers to Pe\~narrubia \& Benson 2005; Mastropietro et al. 2005; Gonz\'alez-Garc\'ia et al. 2005; Aguerri \& Gonz\'alez-Garc\'ia 2009 for interesting discussions on the role played by tidal harassment in the dynamical evolution of satellite galaxies}. 
Our models are purely
gravitational and therefore scale-free. When convenient for
interpretive purposes, we scale our numerical units to physical
parameters similar to those estimated for the Milky Way by Klypin,
Zhao \& Somerville (2002): $V_{\rm max}=188$ km/s and $r_{\rm max}=18$
kpc.

\begin{figure*}   
\plotone{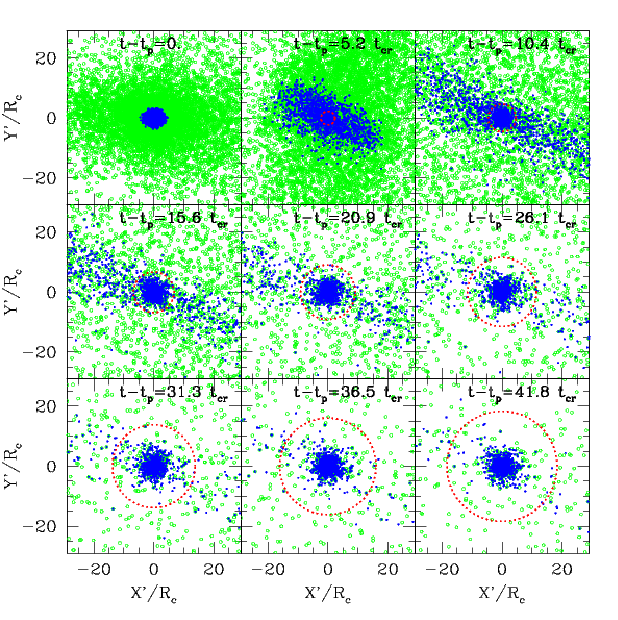}   
\caption{ Time sequence illustrating the evolution of a dSph on a
  highly-eccentric ($r_{\rm peri}$:$r_{\rm apo}=1$:$200$) orbit. Stars
  and dark matter particles are shown in the plane of the orbit and
  are plotted with filled and open circles, respectively, assuming an initial
  stellar segregation of $R_k/r_{\rm max}^d=0.1$. Each panel
  corresponds to a different time, is centered on the bound stellar
  core, and is labeled by the time elapsed since pericenter, $t-t_p$,
  given in units of the crossing time $t_{\rm cr}=R_c/\sigma_0$ of the
  dwarf. Dotted circles indicate the position of the ``break'' radius
  in the density profile discussed in the text.}
\label{fig:xy}   
\end{figure*}   

\begin{figure*}   
\plotone{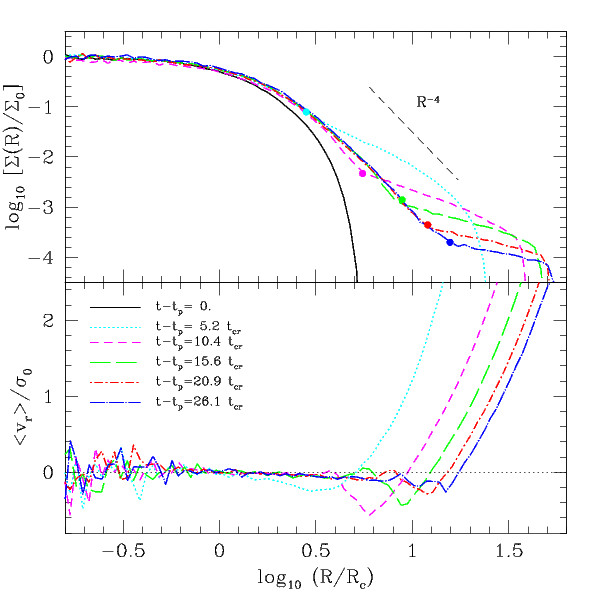}   
\caption{ {\it Upper panel}: Projected stellar density profilesof the
  models shown in Fig.~\ref{fig:xy}. Note how an initial King model
  model (solid line) evolves gradually towards an $R^{-4}$ power-law
  profile (dashed segment) because of tidal stripping. Filled circles
  indicate the ``break'' radius where the profile deviates
  significantly from the equilibrium profile the dwarf approaches at
  late times.  {\it Lower panel}: Average radial velocity of stars
  (measured in 3D spherical shells) in the rest frame of the bound
  stellar core, as a function of radius. This panel shows that the
  excess of stars at large radii is due to particles moving away as
  they escape or settle gradually into their new, more loosely bound
  orbits.}\label{fig:rho}
\end{figure*}   

\subsection{Dwarf galaxy model}\label{sec:dSphmod}
Our dSph models assume that the stellar component follows a King
(1966) sphere embedded within a CDM halo. The dark halo is modeled as
an N-body realization of a spherical NFW profile and is constructed by
generating a $5$ million-particle equilibrium realization using a code
kindly made available by S.Kazantzidis (Kazantzidis et al. 2004,
2006).  Using the same physical scaling adopted for the host galaxy
(\S~\ref{sec:hostmodel}) we assume for the dwarf halo that $r_{\rm
  max}^d=4.4$ kpc and $V^d_{\rm max}=28.6$ km/s. Hereafter we use the
superscripts $d$ and $h$ to identify parameters corresponding to the
dwarf and the host, respectively.
 
Motivated by the large mass-to-light ratio ($M/L$) exhibited by the Milky Way dSphs 
(20$\simless M/L\simless $1000; Mateo 1998; Simon \& Giha 2007) we assume that stars contribute negligibly to the potential, and may therefore be followed by assigning an energy-dependent $M/L$ to each dark matter particle in the equilibrium halo (see
Paper II for details). The outcome is a subset of particles that
follows a King model in equilibrium within the dark halo. The surface
density profile of the initial stellar component can be written as

\begin{eqnarray}   
\Sigma (R)= k \{[1+(R/R_k)^2]^{-1/2} - [1+(R_t/R_k)^2]^{-1/2}\}^2  
\label{eq:Sigma}   
\end{eqnarray}   
where $R$ is the projected radius and $k$ an arbitrary constant.  We
consider models with values for the King ``concentration'' $c_K\equiv
R_t/R_k=5$ and $10$, but note that none of our main conclusions
depends on this choice. For an easier comparison with observations, we
also use the core radius $R_c$, which is estimated independently of
any fitting procedure as the radius where the surface density drops to
one half of its central value. Note that for a King model
$R_c\rightarrow R_k$ as $c_k\rightarrow \infty$. For concentrations
$c_k=5$ and 10, $R_c/R_k=0.84$ and 0.92, respectively.

The main parameter of the model is the degree of spatial segregation
between stars and dark matter in the dwarf, which we may express as
the ratio between the stars' King radius, $R_k$, and $r_{\rm max}^d,
$the radius where the halo circular velocity peaks. The segregation
between stars and dark matter is constrained by the shape of the
stellar velocity dispersion profiles. As discussed in Paper I, only
systems where stars are deeply embedded within the dark halo have
projected velocity dispersion profiles, $\sigma_p(R)$, that, in
agreement with observations, remain nearly {\it flat} almost out to the
``tidal'' radius. For the eight Milky Way dSphs examined in Paper I,
we show that spatial segregations in a broad range, $0.02 \le R_k/r_{\rm
  max}^d\le 0.14$, satisfy this constraint.

From a numerical point of view, the more deeply
segregated the stellar component is, the fewer particles are available
to trace it, and therefore large numbers of particles are required to
resolve the innermost regions of halos where the stars are thought to
reside.  Using a total number of $5\times 10^6$ dark matter particles
for the NFW profile we obtain stellar components traced by $1.95\times
10^4$, $6.86\times 10^4$, and $1.88\times 10^5$ particles, for
$R_k/r_{\rm max}^d=0.02$, $0.05$, and $0.10$, respectively.

\subsection{The Orbits}\label{sec:orbits}

The dwarf galaxy models are placed on highly eccentric orbits of fixed
apocenter, $r_{\rm apo}=10 \, r_{\rm max}=180$ kpc. We ran several
simulations in order to explore the result of varying the pericentric
distance. As expected, the smaller the pericenter of the orbit the
stronger the tidal effects. This may be quantified by the theoretical
``tidal radius'' ($r_{\rm tid}$), defined by (Paper II) 
\begin{equation}
\langle \rho^{\rm d} \rangle (r_{\rm tid})=3\, \langle \rho^{\rm
  h}\rangle (r_{\rm peri}),
\label{eq:rt}
\end{equation}
where $\langle \rho \rangle(r)$ denotes the mean enclosed density
within $r$. Only when $r_{\rm tid}$ is comparable to the size of the
stellar component will the dwarf experience a sizeable change in the
structure of its stellar component. Since we are interested in systems
that {\it do} lose stars to tides after the first pericentric encounter, 
we concentrate here on orbits with
rather small pericenters; i.e., $r_{\rm peri}$:$r_{\rm apo}=$1$:$50$,
1$:$100$ and $1$:$200$, which lead to substantial mass loss.
Table~\ref{tab:mod} summarizes the main parameters of each run. The
table lists the orbital parameters of each orbit, as well as the
fraction of dark mass contained initially within $r_{\rm tid}$. It
also lists the fraction of stars within $r_{\rm tid}$ for our choice
of segregation parameter $R_k/r_{\rm max}^d=0.1$, and two possible
values of the King concentration, $c_K=5$ and $10$. These models are integrated only until their first orbital apocenter. When convenient, we also use some of the dwarf galaxy models introduced in Paper II, which follow the evolution of dwarf galaxies for several orbital periods.

\begin{figure*}   
\plotone{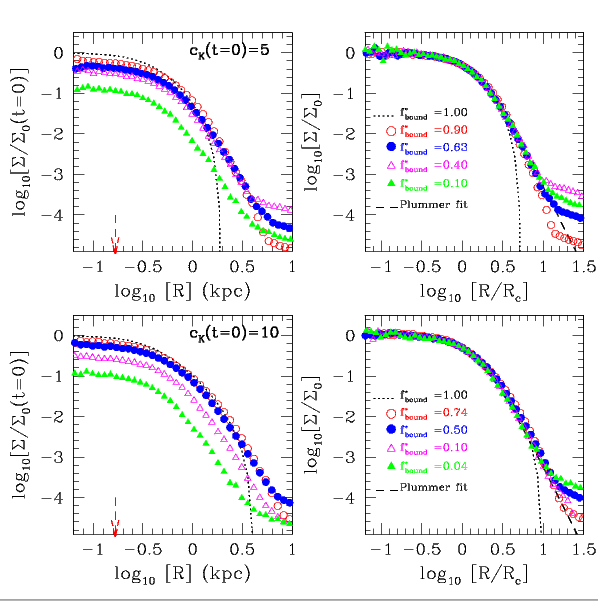}   
\caption{
{\it Left panels}: Stellar surface density profiles of dSphs that have
lost various fractions of their total mass to tides (indicated by the
remaining bound stellar fraction, $f^*_{\rm bound}$.  The initial King
model is shown with black dotted lines. Profiles are normalized to the
{\it initial} central surface density $\Sigma_0(t=0)$. Upper and lower
panels show models with different initial King concentrations $c_K$.
{\it Right panels}: Same profiles as in left, but normalized to the
current values of the central surface density and of the core radius.
The latter is defined as the radius where $\Sigma$ drops to one-half
of its central value.  These panels show that (i) the initial
truncation of the light profile quickly disappears as a result of
tides, and that (ii) the bound stellar remnants evolve towards a
spatial distribution that follows a $\Sigma
\sim R^{-4}$ profile at large radii. }
\label{fig:dens_evol}   
\end{figure*}   

\subsection{The N-body code}
We use {\sc superbox}, a highly efficient particle-mesh code (see Fellhauer et al. 2001 for details) to track the evolution of our dSph N-body models in the host potential. {\sc superbox} uses three nested grid zones with different size centered on the highest-density particle cell of the dwarf. This center is updated every time-step, so that all grids follow the satelly galaxy along its orbit.

Our numerical setup is outlined in detail in Paper II. For completeness we summarize here the main numerical parameters. Each of the three grids has $128^3$ cells. (i) The inner grid resolves the inner region of the dwarf and has a spacing of $dx=r_s^d/126\simeq 8\times10^{-3}r_s^d$. (ii) The middle grid covers the whole dwarf with a spacing that is $dx=r_{\rm vir}^d/126$. (iii) lastly, the outermost grid extends out to $50r_{\rm vir}^d$ and is meant to follow particles that are stripped from the dwarf and orbit within the host galaxy.
{\sc superbox} uses a leap-frog scheme with a constant time-step to integrate the equations of motion, which is selected according to the criterion of Power et al. (2003). Applied to our dwarf galaxy models, this yields $\Delta t=4.6$ Myr. In Paper II we include numerical checks that show that the above selection of parameters leads to N-body dSph models that are in dynamical equilibrium and do not evolve away from the original configuration in isolation. 

To simplify the analysis of the N-body experiments, we define mass, size, velocity and time N-body units as $[M]=5.6\times 10^10 M_\odot$, $[R]=3.5$ kpc, $[V]=262$ km/s and $[T]=0.013$ Gyr. In these units, the gravitational constant is $G=1$.

\section{Results}\label{sec:stellar}
\subsection{Tidal effects on eccentric orbits}

Fig.~\ref{fig:xy} illustrates the evolution of one of our dwarf models
after first pericentric passage on a highly eccentric orbit. Dark halo
 particles and ``stars`` are shown with open and filled circles, respectively,
assuming a star-to-dark matter segregation parameter, $R_k/r_{\rm
  max}^d=0.1$ and $c_K=5$ for the King model representing the
stars. The case shown corresponds to a rather extreme orbit, where the
dwarf is stripped of $\sim 98\%$ of its initial dark mass and $\sim
60\%$ of its stars during the first pericentric passage. Particles are
shown projected on the plane of the orbit, at various times following
pericenter. Each panel is labelled by the time elapsed since
pericenter, $t-t_p$, expressed in units of the crossing time of the
stars at the core radius, as measured by $t_{\rm cr}\equiv
R_c/\sigma_0$, where $\sigma_0$ is the central line-of-sight stellar
velocity dispersion\footnote{To minimize numerical noise, we calculate the central velocity dispersion at $R_c/2$. Since our dwarf models exhibit a flat velocity dispersion profile (\S\ref{sec:dSphmod}) this particular choice of radius has a negligible impact on the value of $\sigma_0$}.  To let the models resettle into dynamical equilibrium, 
we measure these quantities at orbital apocenter.

The panels in Figure~\ref{fig:xy} show that, for such eccentric
orbits, the effect of tides can be well approximated by an impulsive
perturbation acting at pericenter. Indeed, before $t=t_p$ the stellar
system shows almost no sign of being perturbed. A short time after
pericenter, the stars and dark matter that have gained energy during
the interaction start leaving the dwarf through two prominent ``tidal
tails''.

Once enough time has elapsed to allow unbound material to leave the
galaxy, the prominence of the tidal tails wanes and the remaining
bound core of stars resettles into equilibrium. Few obvious ``tidal
features'' remain then. At $t-t_p=41.8\, t_{\rm cr}$ (bottom right
panel in Figure~\ref{fig:xy}) it would be difficult to guess that the
dwarf has lost $60\%$ of its original stars (see Paper II). As
anticipated in Sec.~1, note that almost no dark matter is left
surrounding the bound stellar core.

The process recurs after every pericentric passage. We refer the
interested reader to Paper II for details regarding how the structural
parameters of the dwarf (core radii, velocity dispersion, etc) evolve
after each mass loss episode. We concentrate below on the
tidally-driven evolution of the surface brightness profile of the
stellar component.

\subsection{The equilibrium profile of a tidally-stripped dSph}\label{sec:stellar}

The upper panel of Fig.~\ref{fig:rho} shows the stellar surface
density profile (projected onto the orbital plane) at the various
snapshots shown in Fig.~\ref{fig:xy}. Each curve corresponds to a
different time, after normalizing to the current values of the central
density and of the core radius. The solid curve is the original
undisturbed King profile.

This panel illustrates a few interesting points. The most obvious is
that the stellar profile changes substantially as a result of the
interaction. The outer regions of the profile become distended,
although the shape in the inner regions is much less affected.  As
discussed in Paper II, force-fitting a King-model to the
post-perturbation profile would result in much larger values of the
King concentration, $c_K=R_t/R_k$. It is clear, however, that as a result of stellar stripping the
outer profile loses the well-defined cutoff characteristic of a King
sphere, and approaches instead a power-law at late times (see also Oh et al. 1995). 
This is not
unexpected; the tidal impulse does not only remove some stars by
endowing them with speeds exceeding the local escape velocity, but
also shifts a number of stars to very loosely bound orbits, in effect
populating phase-space all the way to zero energy, $E\sim 0$. A simple
calculation shows that, under these circumstances, a power-law outer
density profile may be expected once the dwarf re-equilibrates (White
1987; Jaffe 1987). The dashed segment shows that $\Sigma \propto
R^{-4}$ describes the outer equilibrium envelope well.

All dSph models that have lost an appreciable fraction of stars to
tides seem to develop an outer power-law profile. This is shown in
Fig.~\ref{fig:dens_evol}, where we show the projected stellar density
profile of several dwarfs after being perturbed by tides. The profiles
are computed at apocenter, after the bound stellar core has relaxed to
equilibrium, in order to minimize the presence of transients in the
outermost region. (We discuss these transients below.)
Profiles on the left panels of Fig.~\ref{fig:dens_evol} are unscaled,
and show how the system changes as mass is lost. (The unperturbed
system is shown by a dotted line.) As discussed in Paper II, mass loss
leads to a drop in the central surface density but leaves the core
radius of the object almost unchanged.  The top and bottom panels of
this figure correspond to dSph models with initial King concentrations
$c_K=5$ and $10$, respectively. Each curve is labeled by $f_{\rm
  bound}^{*}$, the fraction of their initial stellar mass that remains
attached to the bound core.

As anticipated above, the outer regions of the profile become more
distended, and approach a power-law. Scaling each profile to
its central density and to its core radius (right-hand panels in
Fig.~\ref{fig:dens_evol}) shows that a single law reproduces the
equilibrium profiles of all models (provided they have experienced at
least mild mass loss). A simple parametrization is provided by the 
projected Plummer (1911) law, which we write as
\begin{equation}
\Sigma_P(R)=\frac{\Sigma_0}{[1+(R/R_p)^2]^{2}}.
\label{eq:plummer}
\end{equation}
This conclusion seems independent of the initial concentration of the
King sphere and of the actual orbit. Indeed, the profiles shown in
Fig.~\ref{fig:dens_evol} correspond to different orbits and are shown
at different times, having been selected only to span a wide range in
mass loss.

We conclude that the density profiles of relaxed, tidally-stripped
dwarf spheroidals are well approximated by a Plummer law. This seems
to be a necessary (but not sufficient) condition for identifying
systems in eccentric orbits that have lost a significant amount of
mass to stripping. Interestingly, this result implies that those dSphs
where the outer cutoff in the luminosity profile is sharpest are
likely those that have been {\it least affected} by tidal effects.

\begin{figure*}   
\plotone{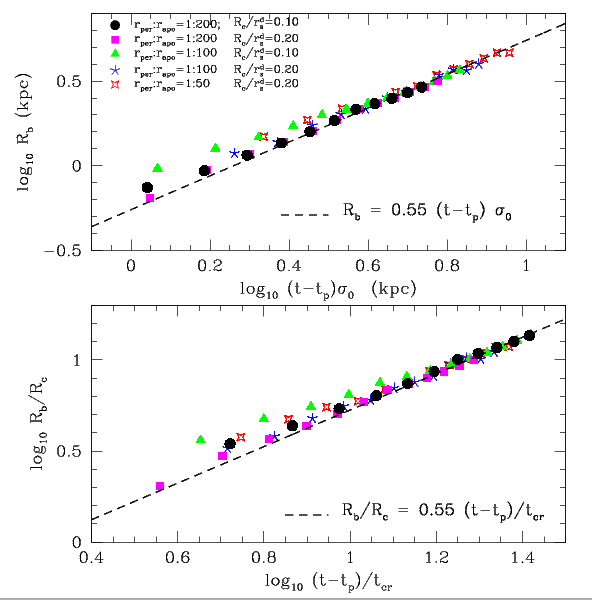}   
\caption{Position of the ``break radius'', $R_b$, measured from the
  sudden upturn in the density profiles (see, e.g., the top panel of
  Fig.~\ref{fig:rho}), as a function of the time elapsed since
  pericenter times the central velocity dispersion $(t-t_p)\sigma_0$
  ({\it upper panel}).  The dotted line shows the result of a simple
  linear fit given by eq.~\ref{eq:rb}. For easier comparison with
  observations, we also show in the bottom panel the same data points
  normalized by the dwarf core radius, $R_c$, and by the
  core crossing time, $t_{cr}\equiv R_c/\sigma_0$. This result provides
  a clock that may be used to assess whether perturbations in the
  outer profiles of dSphs are actually tidal in origin. }
\label{fig:break}   
\end{figure*}   

\begin{figure*}   
\includegraphics[height=180mm, width=180mm]{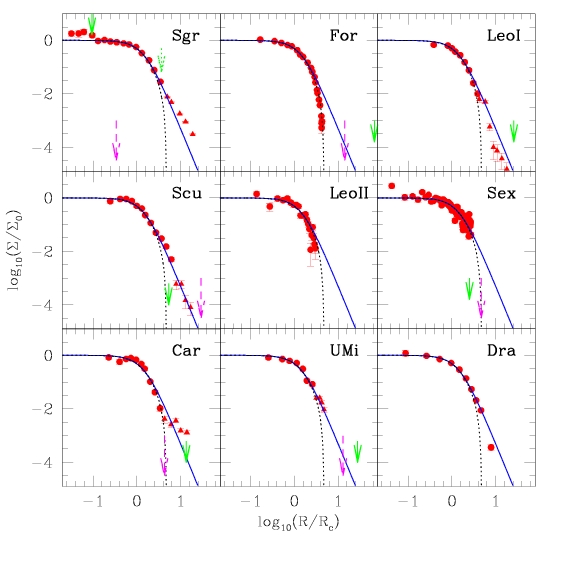}
\caption{
Surface brightness profiles of Sagittarius (from Majewski et
al. 2003), Fornax (Battaglia et al. 2006), Leo I (Sohn et al. 2007),
Sculptor (Westfall et al. 2006), Leo II (Coleman et al. 2007), Sextans
(Irwin \& Hatzidimitriou 1995), Carina (Mu\~noz et al. 2006), UMi
(Palma et al. 2003) and Draco (Martin et al. 2008). All profiles have
been scaled to their central surface density and to the core radius,
defined as the radius where $\Sigma$ drops to one half of its central
value. Solid circles and triangles distinguish, respectively, between
material within and beyond the ``break'' radius {\it reported} in the
literature ($R_{b,{\rm obs}}$). In each panel, the solid line shows a
Plummer law and the dotted line a $c_K=4$ King model.  For systems with available proper motions, solid arrows
indicate the position of the break radius predicted from
eq.~(\ref{eq:rb}) (see also Table~\ref{tab:obs}). For Sagittarius, we
also plot a dotted arrow to show the estimated position of $(R_{b,{\rm
pred}})$ associated to the penultimate pericentric passage. Dashed
arrows denote the position of the tidal radius $r_{\rm tid}$ at
orbital perigalacticon. Only dwarfs where $r_{\rm tid}$ is comparable
or smaller than the luminous radius are expected to be susceptible to
tidal stripping. }
\label{fig:profs}   
\end{figure*}   

\subsection{Tidally-induced transients in the stellar profile}

Another obvious feature in the top panel of Fig.~\ref{fig:rho} is the
presence of a transient ``excess'' in the outer density profile over
the power-law equilibrium profile to which the system relaxes.  The
location where the excess becomes noticeable may be seen as a
``break'' in the outer profile that appears to march outward with
time. The existence of breaks in the light distribution has been often
interpreted as a sign of tidal stripping (Johnston et al. 2002; Mayer
et al. 2002; although see McConnachie et al. 2007 for an alternative
scenario) and may be traced to stars that have gained energy at
pericenter and are still moving away as they settle into their new,
less bound orbits within the dwarf (or escape altogether). This may be
seen in the bottom panel of Fig.~\ref{fig:rho}, which shows the mean
values of the radial velocity component in the dwarf frame as a
function of projected radius for the various snapshots shown in the
top panel. Clearly, the location of the ``break'' in the profile
corresponds roughly to where the average motion of stars transitions
from equilibrium ($\langle v_r \rangle \approx 0$) to becoming
dominated by outward motion ($\langle v_r \rangle > 0$).

As discussed in detail by Aguilar \& White (1986, see also Navarro
1990), the position of the ``break'' correlates with the time elapsed
since pericenter. This provides a ``clock'' that may be used to verify
whether ``breaks'' in observed dSph profiles are actually tidal in
origin. This is shown in Fig.~\ref{fig:break}, where we plot the
position of the break, $R_b$, (which is easily measured from the
density profile; see filled circles in the top panel of
Fig~\ref{fig:rho}) versus the time elapsed since pericenter,
$t-t_p$. The plot shows the results of several dSph models on
different orbits, at various stages of disruption.  The top and bottom
panels of Fig.~\ref{fig:break} show the same data; unscaled in the top
panel and scaled to the core radius and core crossing time in the
bottom panel.  The position of the break radius and the time elapsed
since pericenter are tightly correlated, so that a simple linear fit,
\begin{equation}
R_b= C\, \sigma_0 \, (t-t_p),
\label{eq:rb}
\end{equation}
with $C=0.55\pm 0.03$ approximates well the data in Fig.~\ref{fig:break} (see dashed
line). Although the fit is generally good, it gets better at late
times. Note that at times shortly past pericenter, when the break is
close to the core, eq.~\ref{eq:rb} tends to underestimate the true
radius of the break; by up to $\sim 40\%$ for $R_b\simless 4\, R_c$.

Nevertheless, it is clear that, regardless of the details of the
disruption process, the local crossing time at the location of the
break, $R_b/\sigma_0$, is an extremely good indicator of the time
elapsed since pericenter.  If the orbit and the velocity dispersion of
a dSph are known, then eq.~\ref{eq:rb} offers a simple way of testing
whether a ``break'' seen in the outer profile is tidal in origin.

\section{Application to Local Group Dwarfs}\label{sec:disc}

The results presented above indicate that two conditions must be
satisfied for a ``break'' in the outer profile to be considered tidal
in origin. (i) The dwarf must be susceptible to tides; i.e., the
``tidal radius'', $r_{\rm tid}$ (eq.~\ref{eq:rt}), at perigalacticon
should be comparable or smaller than the luminous radius of the
dwarf. (ii) The break must occur at a radius where the local crossing
time is compatible with the time elapsed since perigalacticon.

The absence of a ``break'' does not, of course, necessarily imply that
a dwarf has not lost stars to tides. Because the ``break'' radius
moves quickly outward, it would be easy to miss as soon as it drifts
into a region where the surface density of stars is low. For the best
studied dwarfs, profiles can be traced robustly out to $\sim 10$ core
radii, where the surface density is typically about $10^{-4}$ times
lower than at the center. The tidal ``break'' in the profiles would
therefore only be detectable for roughly $20$ core crossing times
after pericenter; i.e., for $t-t_p \sim 20 \, R_c/\sigma_0$. This
corresponds to 3 Gyr for a system as bright as Sagittarius, but to
only 300 Myr for a smaller system like Sculptor.  Detecting
unambiguous ``tidal breaks'' thus requires either catching a dwarf
soon after pericentric passage or else deep panoramic surface
photometry to measure robustly the profile at very large radii.

In order to keep the analysis simple, we have used throughout this
analysis profiles onto the plane of the satellite's orbit. This
facilitates the separation between material bound and unbound to the
satellite and sharpens the features in the profile by minimizing the
chance projection of foreground or background unrelated stars and the associated effects
(Read et al. 2006; Mu\~noz et al. 2008). 
Such complication, of course, would need to be considered in detailed
studies of individual objects (e.g. Klimentowski et al. 2007; {\L}okas et al. 2008), 
but is beyond the scope of this paper.


In the case where a break is absent, a tidally-affected dwarf could
still be identified through the presence of a well-defined Plummer
outer law profile. Of course, this is not a sufficient condition (why
couldn't a dSph be born with a Plummer profile?), but it may help to
pinpoint those systems for which further studies might be particularly
fruitful. We caution, however, that distinguishing a Plummer profile
from a King model where the tidal radius is much larger than the core
radius might not be easy. The bottom-right panel of
Fig.~\ref{fig:dens_evol} illustrates this. Distinguishing a $c_K=10$
King model (dotted line) from a Plummer (dashed curve) would require
examining the profile at surface densities below $10^{-4}$ times the
central value. Very few systems have data of such quality. On a more
optimistic note, according to our analysis, dSphs with sharply
truncated outer light profiles are unlikely to have been affected by
tides.

With these caveats in mind, we now apply these lessons to the case of
Milky Way dSphs.  For dwarfs with available proper motions (see
Table~\ref{tab:obs}) we compute their orbits assuming a Klypin et
al. (2002) model for the Milky Way potential.  In our orbit estimates
we neglect in all cases the effect of dynamical friction, since the
mass of these galaxies is very low and we are only interested in the
(short) time elapsed since last perigalacticon. The tidal radius $r_{\rm tid}$ 
of each dSph is calculated using the mass profiles derived in Paper I.
 We begin by
validating our conclusions on the Sagittarius dwarf and discuss next a
few other galaxies with published orbits. We conclude with a
discussion of how these results may apply to the population of
ultra-faint dwarfs recently discovered in the Milky Way halo.

\subsection{The case of Sagittarius}
\label{sec:sgr}

The Sagittarius dwarf (Sgr, for short) offers an ideal testbed for the
ideas presented in the previous section. Sgr is not only the brightest
dSph and one of the closest, but also the one system where evidence
for ongoing tidal stripping is definitive and uncontroversial owing to
the detection of associated tidal streams (see, e.g., Belokurov et
al. 2006b and references therein). Its surface density profile has
been traced to very large radii using 2MASS data (Majewski et
al. 2003) and, because of its proximity ($\simeq 25$ kpc), its orbit
is also fairly well known (Dinescu et al. 2005). Sgr moves on an orbit
that takes it from $\sim 17$ kpc from the Galactic center at
perigalacticon to $\sim 55$ kpc at apogalacticon. It last went through
perigalacticon $\sim 26$ Myr ago and its radial period is $\sim 1.0$
Gyr. The velocity dispersion of Sgr's bound core is $\sigma_0\sim
9.6\pm 1$ km/s (Bellazzini et al. 2008)

According to eq.~\ref{eq:rb}, this implies that Sagittarius should
exhibit a tidal ``break'' at $R_b=0.14$ kpc from the center. Note
that this is well inside Sgr's core radius, which is estimated to be
$\sim 1.5$ kpc.  Interestingly, Majewski et al. (2003) report
the existence a ``bump'' in the stellar profile at about the same
radius (marked by the leftmost arrow in the top-left panel of
Fig.~\ref{fig:profs}). Although the origin of the bump is unclear, our
models suggest that it might be tidal in origin. If true, many stars
around the ``bump'' should be in the process of leaving the dwarf, an
assertion that could in principle be corroborated with velocity data.

The last pericentric passage is thus too recent to have altered the
profile of the Sgr dwarf at large radii. However, the crossing time of Sgr is considerably long ($\simeq 150$ Myr), which suggests that the presence of the ``break'' associated to the {\it penultimate} perigalacticon passage may still be detectable in the outskirts of this galaxy. According to our orbit estimate, the {\it penultimate} perigalacticon passage occurred 1 Gyr ago, so that eq.~\ref{eq:rb} predicts a break radius at 
$R_b\sim 5.4 \pm 2.0$ kpc (dotted arrow in the top
left panel of Fig.~\ref{fig:profs}). This is slightly smaller than the
$R_{b}\sim 7.3$ kpc$\simeq 5 R_c$ break radius claimed by Majewski et
al. (2003) (in this and other panels of Fig.~\ref{fig:profs} we use
filled circles and triangles to report data within/outside the radius
of the break {\it reported} in the literature cited in the
caption). However, the difference is small, considering the
ambiguities in pinning down the exact location of the break
observationally, as well as the fact that eq.~\ref{eq:rb} tends to
{\it underestimate} the location of the break when it occurs as close
to the core as in the case of Sagittarius.


Finally, as a sanity check, the dashed arrow in the Sgr panel of
Fig.~\ref{fig:profs} shows the location of the ``tidal radius''
computed using eq.~\ref{eq:rt} at perigalacticon.  This is only of
order $400$ pc, well inside Sgr's stellar core radius, implying that
Sgr is doomed to lose most of its stellar mass in subsequent orbits.

\subsection{Other Milky Way dSphs}

We analyze here the other five dSphs (aside from Sagittarius) where a
``break'' in the outer profile has been reported in the
literature. Information on these dwarfs (Leo I, Sculptor, Carina, Ursa
Minor, and Fornax) is listed in Table~\ref{tab:obs}. Of the six, only
Leo I has no available measurement of its proper motion, although its
large Galactocentric distance (255 kpc, Caputo et al. 1999; Held et
al. 2001; Bellazzini et al. 2004) together with its extreme
heliocentric radial velocity ($v_{\rm rad, h}\simeq 283$ km/s, Koch et
al. 2007; Mateo 1998, 2008), are suggestive of a highly radial
orbit. We can therefore safely neglect the tangential velocity
component when estimating the time elapsed since the last pericentric
passage.

The surface density profiles of the six dwarfs with claimed ``breaks''
are shown in Fig.~\ref{fig:profs}, together with those of three other
well-studied dSphs (Draco, Leo II and Sextans). Each profile has been
scaled to its individual central surface density and core radius, for
ease of comparison. Solid and dotted lines show, in each panel, a
Plummer and a King ($c_K=4$) model, respectively.  For systems with
full orbital information, we also show the position of the break
radius as predicted from eq.~(\ref{eq:rb}) (solid arrows) and the
tidal radius, $r_{\rm tid}$ (eq.~\ref{eq:rt}), calculated at
perigalacticon (dashed arrows).

\begin{figure*}   
\plotone{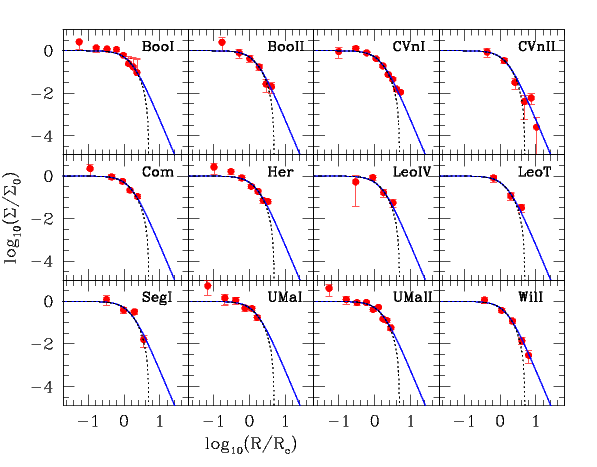}   
\caption{
As Fig.~\ref{fig:profs}, but for 12 of the recently-discovered,
ultra-faint dSphs, as given by Martin et al. (2008). Each profile has
been scaled to the central surface brightness and the core
radius. Solid and dotted lines show the Plummer and King ($c_K=4$)
projected density profiles.}
\label{fig:faint_profs}   
\end{figure*}   

\subsubsection{Fornax}

The Fornax dwarf provides probably the clearest case of a dwarf whose
stellar component has {\it not} been disturbed by Galactic tides. Not
only is the tidal radius clearly well outside the luminous radius, but
also there is no evidence for distensions in the outer profile that
may be ascribed to the past effect of tides. A $c_k=4$ King model fits
the surface density profile of Fornax extremely well, down to the last
measured point.

\subsubsection{Leo I}

The case of Leo I is less clear-cut. Sohn et al. (2007) report a
``break'' in the outer regions, which we highlight with triangles in
the corresponding panel of Fig.~\ref{fig:profs}.  This figure shows
that there is indeed an excess relative to a King model fit to the
inner profile, but not relative to a Plummer law. Application of
eq.~\ref{eq:rb} suggests that a tidal break should be located well
beyond $10 R_c$, and thus outside the region surveyed around this
dwarf. This implies that the outer profile of Leo I should be in dynamical
equilibrium, and that no transients of tidal origin should be
detectable in the data presently available.

This does not exclude the possibility that Leo I might have been
tidally disturbed in the past. Indeed, its outer profile resembles
closely the Plummer profile which, according to our results, tidally
disturbed systems should evolve to. Interestingly, Mateo et al. (2008)
find an increase of the orbital anisotropy for the stellar population
located beyond the reported ``break'' radius.  Further data that
extend out into the region where local crossing times are comparable
to the time elapsed since pericenter are needed to determine whether
this dwarf has really been affected by Galactic tides.

\subsubsection{Sculptor}

Sculptor is also a difficult case. Its projected density profile is
well described by a Plummer law, as would be expected for a relaxed
system that has undergone tidal stripping, but the pericenter of its
orbit seems too large for tides to have played a role. Indeed, we find
that the tidal radius lies well beyond the stellar radius (assuming
the orbital pericenter to be $r_{\rm peri}\sim 76$ kpc, see Table
~\ref{tab:obs}). Eq.~\ref{eq:rb} predicts a break at $\simeq 1$ kpc
(indicated by the solid arrow in Fig.~\ref{fig:profs}), but there is
no obvious feature in the profile at that radius. We conclude that the
``break'' claimed by (Westfall et al. 2006) at $\approx 1.5$ kpc is
{\it not} tidal in origin but likely a result of fitting a King model
to a profile that is best described by a power-law at large radii.

\subsubsection{Carina}

The presence of a break in the outer profile is probably clearest in
the case of the Carina dSph (Majewski et al. 2005). 
As shown in Fig.~\ref{fig:profs},
the surface density profile of this galaxy can be accurately fitted by
a King model with concentration $c_K=4$ inside $R_{b}=0.70$ kpc
$\approx 4 R_c$. At that radius, the light profile shows a clear
inflection point and the profile becomes much shallower out to the
outermost radius surveyed. Mu\~noz et al. (2007) argue that this
feature in the outer profile is tidal in origin.

Our results seem to disagree with this conclusion. According to our
orbit estimate, the tidal radius of Carina at perigalacticon is
roughly $\sim 4 R_c$, which means that Galactic tides could, in
principle, have been important in this galaxy. However,
eq.~\ref{eq:rb} indicates that the break should occur a little more
than twice farther from the center than observed (see solid arrow in
the Carina panel of Fig.~\ref{fig:profs}). This would seem to suggest
that the feature observed in Carina is not tidal in origin.

This conclusion is subject to a few caveats. Our analysis neglects
projection effects that may result in the superposition of unbound
stars onto the main core of the dwarf, as would be the case if we were
seeing a tidal tail ``end on''. This may shift the break radius
inwards, although by how much is not clear at this point. Also, it
should be noted that measuring proper motions for stars at such
distance ($\simeq 100$ kpc) is an extremely difficult task. The
possibility remains, therefore, that the orbit we assume for Carina
(see Table~\ref{tab:obs}) may be in error.  Should a revised orbit in
the future lead to better agreement between the predictions of
eq.~\ref{eq:rb} and the observed location of the break, it would
certainly lend support to the idea that this feature is tidal in
origin and that Carina is in the process of shedding many of its stars
into the Galactic halo.

\subsubsection{Ursa Minor}

The orbit of the Ursa Minor dSph implies a tidal radius well outside
the luminous radius of the dwarf, casting doubt on a tidal
interpretation for the ``break'' at $\sim 0.65$ kpc $\sim 3\, R_c$
claimed by Mart\'inez-Delgado et al. (2001) and Palma et
al. (2003). This conclusion is supported by the fact that the
predicted location of the tidal break, according to eq.~\ref{eq:rb},
would be at roughly $\sim 7$ kpc, or $\sim 20\, R_c$, well beyond the
region surveyed by available data. Unless the orbit for Ursa Minor is
substantially in error, we conclude that the features in the outer
profile of this dwarf are {\it not} tidal in origin.

\subsubsection{Sextans}
As in the case of the UMi dSph, the orbit of Sextans implies a tidal radius that falls beyond the region occupied by stars, which suggests that stellar stripping have not affected this galaxy. Interestingly, the irregular, patchy shape of Sextans (Irwin \& Hatzidimitriou 1995) has been commonly explained as a possible signature of tidal distortions. However, the proper motions recently estimated by Walker et al. (2008) place the orbital pericentre of Sextans at $\approx 70$ kpc and therefore seem to contradict this scenario. On the other hand, the density profile of Sextans is only known out to $\sim 4R_c$ from the galaxy center, which proves insufficient to determine whether or not the luminous profile is truncated at large radii.

\subsection{Dwarfs with unknown orbits}

Of all the systems shown in Fig.~\ref{fig:profs}, Leo II and
Draco do not have published proper motions, so it is not possible to
estimate robustly the time elapsed since perigalacticon. Furthermore,
the radial extent of the data available for Sextans ($R<3\, R_c$) is
too small to allow for conclusive analysis. Leo II appears to be
sharply truncated in the outer regions, but again the radial extent of
the data is probably not good enough to warrant a more detailed
assessment. Draco, on the other hand, has an outer profile that
resembles a Plummer law, so it is possible in principle that tides may
have played a role in the evolution of this system. A determination of
the orbit of these systems, together with data that extend the profile
to larger radii, are needed in order to assess the role of tides in
their evolution.

\subsubsection{Ultra-faint dwarfs}

Fig.~\ref{fig:faint_profs} shows the surface density profiles of $12$
ultra-faint dwarfs recently discovered in SDSS data, as computed by
Martin et al. (2008). The scaling in each profile is the same as in
Fig.~\ref{fig:profs}. Available data for the ultra-faint dwarfs
clearly do not extend sufficiently far in order to allow for the same
kind of analysis we carried out in the previous subsections for their
brighter counterparts. Nevertheless, it is interesting that those
galaxies with deep photometric data, e.g. CVn II and Willman I, seem to
follow a Plummer profile in the outer regions. Although this is not a
sufficient condition to guarantee that these systems have been
strongly perturbed by tides, it implies that the possibility that at
least some of these systems may have been shaped by tides should not
be dismissed out of hand.

\section{summary}\label{sec:sum}

We have used N-body simulations to identify signatures of ongoing or
past tidal stripping that may be used to guide the interpretation of
the observed surface density profile of dwarf spheroidals. Our models
assume cosmologically motivated initial conditions, where the stellar
component of a dSph is represented by a King sphere embedded within an
NFW halo. The modeling assumes that dSphs are on eccentric orbits, so
that tides operate over short periods of time, when the dwarf is at
its orbital perigalacticon.

Our main findings may be summarized as follows.

\begin{itemize}

\item
Only systems in orbits where the tidal radius (eq.~\ref{eq:rt})
(measured at perigalacticon) is comparable to or smaller than the
luminous radius of the dwarf are significantly affected by tides.

\item 
The luminous profile of dwarfs that have undergone at least mild tidal
mass loss is modified substantially, and approaches a Plummer law as
the system gradually resettles into equilibrium after each pericentric
passage.

\item
As the system re-equilibrates, the outer profile shows transient
``bumps'' or ``breaks'' at radii where the local crossing time exceeds
the time elapsed since perigalacticon. For systems with known orbits,
this provides a way of assessing whether breaks in the observed dSph
profiles are actually tidal in origin.

\end{itemize}

Applied to the Sagittarius dwarf (where tidal stripping is beyond
doubt), these results identify two features in the surface brightness
profile that may be traced to its two last pericentric
passages. Encouraged by this, we have also checked the plausibility of
a tidal origin for the ``breaks'' reported in other dSphs.

We are able to rule out a possible tidal origin for the break reported
in Leo I.  Indeed, our results predict that any tidal break would
occur at radii beyond those surveyed by current data. For Carina, our
model indicates that the tidal break should occur at a radius twice as
far as observed.  We conclude that either the outer excess of stars in
Carina is not tidal in origin or that the published proper motions 
for this system are in error.

Sculptor is an intriguing case; its profile matches well a Plummer law
(suggesting that past tidal mass loss could have occurred), but the
tidal radius seems too small for tides to have been important in this
system. A similar comment applies to Draco, although there is no
published orbit for this system, which hinders further
analysis. Available data for Ursa Minor and Sextans do not probe the density
profile far enough out to allow for meaningful constraints. We note,
however, that the tidal radius for both systems are too large for tides
to have been important during their evolution.

Published profiles for other Milky Way dSph companions do not probe
sufficiently far to allow for conclusive assessment. Photometric
surveys that extend surface brightness profiles beyond $\sim 10$ core
radii, together with improved constraints on the orbital parameters of
dSphs, are needed in order to establish whether Galactic tides
actually play an active role in shaping the structure of our faintest
and closest extragalactic neighbours.

\vskip0.5cm JP acknowledges useful comments from M. Wilkinson, M. Bellazzini and the anonimous referee. 
JP and JFN would like to acknowledge the hospitality of
the KITP at the University of California, Santa Barbara where part of
this work was completed. This research was supported in part by the
National Science Foundation under Grant No. NSF PHY05-51164.  AWM
acknowledges support from a Research Fellowship from the Royal
Commission for the Exhibition of 1851 during much of this work. He
also thanks Sara Ellison for additional financial assistance during
this time.

{}

\begin{table}   
\begin{center}  
\caption{Orbital parameters of our models}  
\begin{tabular}{l  c  c c c c c c c} \hline \hline   
$r_{\rm peri}$:$r_{\rm apo}$ & $r_{\rm peri}$ & $T_r$ & $r_{\rm tid}$ & $M^d(<r_{\rm tid})^\P$ & $f^d_{\rm bound}$$^\dagger$ & $f_{\rm bound}^\star$$^\ddagger$ & $f_{\rm bound}^\star$$^\ddagger$ \\   
 & [kpc] & [Gyr] & [$r_s^d$] & [$M^d_{\rm tot}$] &  & ($c_k=5)$ & ($c_k=10)$\\   
\hline  
1:50  & 3.6    & 2.94         & 0.083  & $1.35\times 10^{-3}$ & $6.6\times 10^{-2}$ & 0.91 & 0.50\\  
1:100 & 1.8    & 2.80         & 0.036  & $2.75\times 10^{-4}$ & $3.7\times 10^{-2}$ & 0.65 & 0.30\\  
1:200 & 0.9    & 2.22         & 0.017  & $6.50\times 10^{-5}$ & $2.0\times 10^{-2}$ & 0.38 & 0.17\\  
\hline    
\end{tabular}
\label{tab:mod}  
\end{center} 
\begin{center}
\tablenotemark{$^\P$estimated for an unperturbed model at the pericentric distance}\\ 
\tablenotemark{$^\dagger$calculated at the apocentre following the first pericentric passage}\\ 
\tablenotemark{$^\ddagger$Segregation $R_k/r_{\rm max}^d=0.1$}
\end{center} 
\end{table}

\begin{table*}   
\begin{center}   
\caption{Orbital Parameters of Local Group Satellites with known orbit}  
\begin{tabular}{l c c c c c c c c} \hline \hline   
Galaxy & 
Sgr (last) &
Sgr (penultim.) &
Leo I &
Sculptor &
Carina &
Ursa Minor &
Fornax &
Sextans \\ \hline
$D$ (kpc) &
$25\pm 2$ &
$25\pm 2$ &
$250\pm 30$  &
$87 \pm 4$ &
$100 \pm 5$ &
$76 \pm 3$ &
$138 \pm 8$ &
$86 \pm 4 $\\
$l (^\circ)$ &
6.7 &
6.7&
226&
283 &
260&
105&
237 &
243\\
$b (^\circ)$ &
$-24.4$ &
$-24.4$ &
+49.0&
-84.5&
-22.2&
+44.8&
-65.8&
+43 \\
$v_{r,h}$ (km/s)&
+137&
+137&
+259&
+110&
+223&
-247&
+53 &
+227\\
$\mu_l$ (mas/year)&
$-2.58\pm 0.20$ &
$-2.58\pm 0.20$ &
-- &
$-0.09\pm 0.14$ &
$-0.08\pm 0.09$ &
$0.51\pm 0.17$ &
$0.29\pm 0.05$ &
$0.24 \pm 0.42$ \\
$\mu_b$ (mas/year)&
$2.07\pm 0.20$ &
$2.07\pm 0.20$ &
-- &
$0.01\pm 0.13$ &
$0.25\pm 0.09$ &
$0.20\pm 0.17$ &
$0.52\pm 0.05$ &
$-0.14\pm 0.42$ \\
$R_c$ (kpc) &
1.50&
1.50&
0.17&
0.1&
0.18&
0.20&
0.40&
0.32\\
$\sigma_0$ (km/s) &
$9.6 \pm 1.0$ &
$9.6 \pm 1.0$ &
$8.8 \pm 1.3$ &
$6.6 \pm 1.0$ &
$6.8 \pm 1.0$ &
$9.3 \pm 2.0$ &
$11.1 \pm 0.7$&
$6.6\pm 0.7 $   \\
$r_{\rm peri}$ (kpc) &
$16.5 \pm 2.4$ &
$16.5 \pm 2.4$ &
0 &
$76 \pm 4$ &
$24 \pm 5$ &
$48 \pm 8$ &
$129 \pm 17$&
$69 \pm 9$  \\
$t-t_p$ (Gyr) &
$0.023 \pm 0.011$ &
$1.0 \pm 0.2$ &
$1.02\pm 0.12$ &
$0.28 \pm 0.03$ &
$0.81 \pm 0.12$ &
$1.5 \pm 0.3$ &
$6.3 \pm 2.9$ &
$0.32 \pm 0.08$ \\
$R_{b,{\rm obs}}$ (kpc) &
$7.3 \pm 0.6$ &
$7.3 \pm 0.6$ &
$0.58\pm 0.14$ &
$1.3 \pm 0.2$ &
$0.70 \pm 0.14$ &
$0.65 \pm 0.10$ &
$2.4 \pm 0.2$   &
-- \\
$R_{b,{\rm pred}}$ (kpc) &
$0.14 \pm 0.08$ &
$5.4 \pm 2.$ &
$5.1\pm 1.3$ &
$1.04 \pm 0.11$ &
$3.1 \pm 0.4$ &
$7.9 \pm 1.5$ &
$39 \pm 18$ &
$1.2 \pm 0.4$\\
refs. &
 1,2,3 &
 1,2,3 &
4,5,6 &
7,8,9 &
5, 10,11,12&
5,13,14,15&
5,16,17 &
5, 18\\
\hline    
\newline   
\newline   
\end{tabular}\label{tab:obs}   
\tablerefs{(1) Majewski et al. (2003); (2) Dinescu et al. (2005); (3) Bellazzini et al. (2008);
(4) Caputo et al. (1999); (5) Mateo (1998, 2008); (6) Koch et al. (2007);
(7) Walcher et al. (2003); (8) Westfall et al. (2006); (9) Piatek et al. (2006);
(10) Majewski et al. (2000, 2005); (11) Mu\~noz et al. (2006); (12) Piatek et al. (2003); 
(13) Palma et al. (2003); (14) Mu\~noz et al. (2005); (15) Piatek et al. (2005);
(16) Rizzi et al. (2007); (17) Piatek et al. (2007); (18) Walker et al. (2008)
}  
\end{center}   
\end{table*}

\end{document}